\documentstyle[preprint,prl,aps]{revtex}
\begin{document}
\draft

\title{Bose-Einstein Condensation and Intermediate State of the
Photon Gas}

\author{ Levan N. Tsintsadze}
\address{Venture Business Laboratory, Hiroshima University, Higashi-Hiroshima,
Japan}

\date{\today}

\maketitle

\begin{abstract}

Possibility of establishment of equilibrium between the photon and
the dense photon bunch is studied. In the case, when the density
of plasma does not change, the condition of production of the
Bose-Einstein condensate is obtained. It is shown that the
inhomogeneity of density of photons leads to a new intermediate
state of the photon gas.

\end{abstract}


Importance of scattering processes in interaction of photons with
an electron gas was noted for the first time by Kompaneets
\cite{kom}. Who has shown that the establishment of equilibrium
between the photons and the electrons is possible through the
Compton effect. The role of the Compton effect in setting
equilibrium between the photons and the electrons was discussed by
Kompaneets in the nonrelativistic approximation. In his
consideration, since the free electron does not absorb and emit,
but only scatters the photon, the total number of photons is
conserved. Taking into account the fact that for the high
frequencies of photons the probability of Compton scattering
exceeds the absorption probability, Kompaneets has derived the
kinetic equation for the photon distribution function. Using the
kinetic equation of Kompaneets, Zel'dovich and Levich \cite{zel}
have shown that in the absence of absorption the photons undergo
Bose-Einstein condensation. Such a possibility for the
Bose-Einstein condensation occurs in the case, when the processes
of change of energy and momentum in scattering dominates over the
processes involving change of photon number in their emission and
absorption. Thus, in the above papers \cite{kom},\cite{zel} the
authors demonstrated that the usual Compton scattering leads to
the establishment of equilibrium state and Bose-Einstein
condensation in the photon-plasma medium.

In the present letter, we show that exists an another new
mechanism of creation of equilibrium state and Bose-Einstein
condensation in a nonideal dense photon gas. In our consideration
we assume that the intensity of radiation (strong and super-strong
laser pulse, non-equilibrium cosmic field radiation, etc.) is
sufficiently large, so that the photon-photon interaction can
become more likely than the photon-electron interaction. Under
these conditions in the case of relativistic intensities of
radiation of electromagnetic (EM) waves, we may consider the
system of two weakly interacting subsystems: the photon gas and
the plasma, which slowly exchange energy between each other. In
other words, relaxation in the photon-plasma system is then a
two-stage process: firstly the local equilibrium is established in
each subsystem independently, corresponding to some average
energies $E_\gamma=K_BT_\gamma$ and $E_p=K_BT_p$, where $K_B$ is
the Boltzmann constant, $T_\gamma$ is the characteristic
"temperature" of the photon gas, and $T_p$ is the plasma
temperature.

As it was mentioned above, in the case of the strong radiation of
EM waves there is an appreciable probability that the
photon-photon interaction takes place, the phases of the waves
will, in general, be random functions of time. We need therefore
not be interested in the phases and can average over them. Under
these conditions the perturbation state of the photon gas can be
described in terms of the occupation number
$N(\vec{k},\omega,\vec{r},t)$ of photons and study how these
numbers change due to the processes of interaction of the photons
with others or with plasma electrons. Here,
$N(\vec{k},\omega,\vec{r},t)$ is the slowly varying function in
space and time.

A new version of kinetic equation for the occupation number
$N(\vec{k},\omega,\vec{r},t)$ of photons, for modes propagating
with wavevector $\vec{k}$ and frequency $\omega$, at a position
$\vec{r}$ and time t, was derived in
Refs.\cite{ltsin98},\cite{ntsin98},\cite{men}
\begin{eqnarray}
\frac{\partial}{\partial
t}N(\vec{k},\omega,\vec{r},t)+\frac{c^2}{\omega}(\vec{k}\cdot\nabla)
N(\vec{k},\omega,\vec{r},t)
-\omega_p^2sin\frac{1}{2}\Bigl(\nabla_{\vec{r}}\cdot
\nabla_{\vec{k}}-\frac{\partial}{\partial
t}\cdot\frac{\partial}{\partial\omega}\Bigr)\rho\frac{N}{\omega}=0
\ , \label{kin}
\end{eqnarray}
where $\omega_p=\sqrt{\frac{4\pi e^2n_0}{m_{0e}}}$, $m_{0e}$ is
the electron rest mass, n and $n_0$ are the non-equilibrium and
equilibrium densities of the electrons, respectively,
$\rho=\frac{n}{n_0}\frac{1}{\gamma}$, and $\gamma$ is the
relativistic gamma factor of the electrons and can be expressed as
\begin{eqnarray}
\gamma=\sqrt{1+Q}=\sqrt{1+\beta\int\frac{d\vec{k}}{(2\pi)^3}\int\frac{d\omega}{2\pi}\frac{
N(\vec{k},\omega,\vec{r},t)}{\omega}} \ , \label{gam}
\end{eqnarray}
where $\beta=\frac{2\hbar\omega_p^2}{m_{0e}n_0c^2}$ and $\hbar$ is
the Planck constant divided by $2\pi$.

Equation (\ref{kin}) is the generalization of the Wigner-Moyal
equation for the classical EM field. Note that from this equation
follows conservation of the total number of photons, but not the
momentum and energy of photons, i.e.,
\begin{eqnarray}
N=2\int
d\vec{r}\int\frac{d\vec{k}}{(2\pi)^3}\int\frac{d\omega}{2\pi}
N(\vec{k},\omega,\vec{r},t)=const. \ , \label{con}
\end{eqnarray}
where coefficient 2 denotes two  possible polarization of the
photons. Hence, the chemical potential of the photon gas is not
zero.

In the geometric optics approximation ($sinx\approx x$)
Eq.(\ref{kin}) reduces to the one-particle Liouville-Vlasov
equation
\begin{eqnarray}
\frac{\partial}{\partial
t}N(\vec{k},\omega,\vec{r},t)+\frac{c^2}{\omega}(\vec{k}\cdot\nabla)
N(\vec{k},\omega,\vec{r},t)
-\frac{\omega_p^2}{2}\Bigl(\nabla_{\vec{r}}\rho\cdot\nabla_{\vec{k}}-\frac{
\partial\rho}{\partial
t}\cdot\frac{\partial}{\partial\omega}\Bigr)\frac{N(\vec{k},\omega,\vec{r},t)}{\omega}=0
\ . \label{liou}
\end{eqnarray}
We specifically note here that in Eqs. (\ref{kin}) and
(\ref{liou}) there are two forces of distinctive nature, which can
change the occupation number of photons. One force appears due to
the redistribution of electrons in space, $\nabla n_e$, and time,
$\frac{\partial n_e}{\partial t}$. The other force arises by the
variation of shape of wavepacket. In other words, this force
originates from the alteration of the average kinetic energy of
the electron oscillating in the rapidly varying field of the EM
waves ($\nabla\gamma,$ $\frac{\partial\gamma}{\partial t}$).

In the following we shall consider the case, when the slowly
varying function $N(\vec{k},\omega,\vec{r},t)$ essentially changes
on the distance L and in time t, which satisfy the inequalities
$\omega_pt\gg 1$ and $L\gg\frac{c}{\omega_p}$. Under these
conditions, the density of plasma almost does not change in
comparison with the energy of EM field, i.e., we take into account
the forces only due to the variation of $\gamma$. In this
approximation, the statistical equilibrium between photons and the
photon bunch will establish itself as a result of the "Compton"
scattering processes.

For convenience we introduce momentum and the "rest" energy of the
photon as $\vec{p}=\hbar\vec{k}$ and $m_\gamma
c^2=\frac{\hbar\omega_p}{\gamma^{1/2}}$, respectively. With these
notations in the quasi-stationary case Eq.(\ref{liou}) takes the
form
\begin{eqnarray}
\vec{p}\cdot\nabla_{\vec{r}}
N(\vec{p},\vec{r})-\frac{1}{2}\nabla_{\vec{r}} (m_\gamma
c)^2\cdot\nabla_{\vec{p}} N(\vec{p},\vec{r})=0 . \label{quasi}
\end{eqnarray}

The photon gas is characterized by an arbitrary initial
distribution with some average energy. This may, in particular, be
a Bose-Einstein distribution with a characteristic temperature
$T_\gamma$, which, in general, will differ from the plasma
temperature. Slower processes of the equalization of the photon
and the plasma particle temperatures will take place afterwards.

The solution of Eq.(\ref{quasi}) can be written as
\begin{eqnarray}
N(\vec{p},\vec{r})=\Bigl(exp\Bigl\{\frac{c\sqrt{p^2+m_\gamma^2c^2}-\mu_\gamma}{
T_\gamma}\Bigr\}-1\Bigr)^{-1} \ , \label{sol}
\end{eqnarray}
where $\mu_\gamma$ is the chemical potential of the photon gas.

From Eq.(\ref{sol}) follows that the energy of the photon may be
written as
\begin{eqnarray}
\varepsilon_\gamma=c\sqrt{p^2+m_\gamma^2c^2} \ . \label{ener}
\end{eqnarray}
The form of expression (\ref{ener}) coincides with the expression
for the relativistic energy of material particles, so that the
rest mass $m_\gamma$ is associated with the photon in a plasma. We
note that in contrast to material particles, the rest mass of
photons depends on the plasma density and the relativistic gamma
factor through the electron mass. Hence, the rest mass of photons
is the function of the spatial coordinate $\vec{r}$.

The photon density $n_\gamma$, chemical potential, temperature and
relativistic gamma factor are related by the expression
\begin{eqnarray}
n_\gamma=2\int\frac{d\vec{p}}{(2\pi\hbar)^3}N(\vec{p},\vec{r})=2\int
\frac{d\vec{p}}{(2\pi\hbar)^3}\frac{1}{z^{-1}e^{\frac{\varepsilon_k}{
T_\gamma}}-1} \ ,\label{relat}
\end{eqnarray}
where $z=exp\Bigl\{-\frac{m_\gamma
c^2-\mu_\gamma}{T_\gamma}\Bigr\}$ is the fugacity, and
$\varepsilon_k=c\sqrt{p^2+m_\gamma^2c^2}-m_\gamma c^2$ is the
kinetic energy of the photon.

Requirement that the photon density must be positive leads to the
important relation $z^{-1}e^{\frac{\varepsilon_k}{ T_\gamma}}>1$,
which should be satisfied for any energies, $\varepsilon_k$,
including the ground state, $\varepsilon_k=0$, and at any points
of space, i.e., the following inequality must hold
\begin{eqnarray}
m_\gamma(\vec{r})c^2=\frac{\hbar\omega_p}{\gamma^{1/2}(\vec{r})}>\mu_\gamma
\ . \label{mgmu}
\end{eqnarray}
The critical temperature of the Bose-Einstein condensation is
determined for the fixed points, $r_f$, from the condition
\begin{eqnarray}
m_\gamma(r_f)c^2=\frac{\hbar\omega_p}{\gamma^{1/2}(r_f)}
=\mu_\gamma \ . \label{rf}
\end{eqnarray}

After integration over the momentum, with condition
$\mu_\gamma=m_\gamma c^2$, from Eq.(\ref{relat}) we obtain
\begin{eqnarray}
n_\gamma=\frac{1}{\pi^2\lambda_c^3}\frac{1}{a}\sum_{\eta=1}^\infty\frac{e^{\eta
a}}{\eta}K_2(\eta a) \ , \label{ngam}
\end{eqnarray}
where $\lambda_c=\frac{\hbar}{m_\gamma
c}=\frac{c}{\omega_p}\gamma^{1/2}$ is the Compton wavelength of
photon, $a=\frac{m_\gamma c^2}{T_\gamma}$, $K_2(X)$ is the
McDonald function of second order, and $\eta$ is an integer.

We now first consider the non-relativistic "temperatures", i.e.,
when $a\gg 1$. Since, in this case
$K_2(X)\simeq\Bigl(\frac{\pi}{2X}\Bigr)^{1/2}e^{-X}$, we get the
following expression for the critical temperature from Eq.
(\ref{ngam})
\begin{eqnarray}
T_c=\frac{2^{1/3}\pi\hbar^2}{\zeta^{2/3}(3/2)m_\gamma}n_\gamma^{2/3}\
, \label{Tcnon}
\end{eqnarray}
where $\zeta(3/2)=2.612$ is the Rieman zeta function.

For the ultrarelativistic photon gas, when the characteristic
temperature satisfies the inequality $m_{0e}c^2\gg T_\gamma\gg
m_\gamma c^2$, for the critical temperature we obtain
\begin{eqnarray}
T_c=\hbar c\Bigl(\frac{\pi^2}{\zeta(3)}\Bigr)^{1/3}n_\gamma^{1/3}
\ . \label{Tcultra}
\end{eqnarray}

In the case, when the temperature $T_\gamma$ is below the critical
temperature $T_c$ of the Bose-Einstein condensation, the
occupation number can be written in the form
\begin{eqnarray}
N(\vec{p},\vec{r})=\frac{1}{exp\Bigl(\frac{c\sqrt{p^2+m_\gamma^2c^2}-m_\gamma
c^2}{T_\gamma}\Bigr)-1}+4\pi^3n_{0\gamma}\delta(\vec{p})\ .
\label{NTlow}
\end{eqnarray}
Here, the second term corresponds to the finite probability for
the photon to have exactly zero momentum, and $n_{0\gamma}$ is the
density of photons in the ground state.

Using Eq.(\ref{NTlow}), we now determine density of the photons
that fall into Bose-Einstein condensate. First, in the
non-relativistic limit ($m_\gamma c^2>T_\gamma$), we have
\begin{eqnarray}
n_{0\gamma}=n_\gamma\Bigl(1-\Bigl(\frac{T_\gamma}{T_c}\Bigr)^{3/2}\Bigr)\
, \hspace{2cm} T_\gamma < T_c\ . \label{Nogam}
\end{eqnarray}
Next, for the ultrarelativistic photon gas ($m_{0e}c^2 > T_\gamma
> m_\gamma c^2$), we get
\begin{eqnarray}
n_{0\gamma}=n_\gamma\Bigl(1-\Bigl(\frac{T_\gamma}{T_c}\Bigr)^{3}\Bigr)\
, \hspace{2cm} T_\gamma < T_c\ . \label{Nogamrel}
\end{eqnarray}
It should be noted that the production of the Bose-Einstein
condensates is realized, as follows from Eqs. (\ref{mgmu}) and
(\ref{rf}), in regions, where the field intensity is large.

Now we will show that the inhomogeneity of intensity of the EM
field leads to the intermediate state of the Bose-Einstein
condensates of the photon gas. The point is that the relativistic
gamma factor, $\gamma$, is the function of spatial coordinates and
at different points $\gamma$ is different, which means that the
effective mass $m_\gamma$ of the photon, as a function of
$\gamma$, also changes in the usual space. Hence, one can find
some points or regions of high intensities in the gas volume,
where equality (\ref{rf}) is fulfilled, so that the photon gas is
found in the Bose-Einstein condensate state, whereas in the rest
of the domains or points the photons remain in normal state, for
which inequality (\ref{mgmu}) is satisfied. We can therefore
conclude that in such a case the photon gas, as whole is in the
intermediate state. This effect resembles the Landau's
intermediate state in superconductor.

In order to make this statement more clear, we consider
periodically modulated cylindrically shaped laser pulse (light
bullet). We assume that the field intensity is uniform in the
transverse direction (in general, it is not necessary) of
propagation of pulses and periodic along the light bullet (1D
optical lattices). In this case the relativistic gamma factor has
the following form
\begin{eqnarray}
\gamma=\sqrt{1+Q_0(1+dcosqz)} \ ,\label{glb}
\end{eqnarray}
where $Q_0$ is a constant, d  and $\frac{2\pi}{q}$ are the depth
and the period of modulation, respectively.

Furthermore, we assume the periodic boundary conditions
$N(\vec{p},z)=N(\vec{p},z+L)$, and $\gamma (z)=\gamma (z+L)$, with
L being the length of the light bullet.

We then suppose that at all points $z=\frac{2\pi}{q}l$ (where
$l=0,1,2,3,...$) photons are in the Bose-Einstein condensate
state, that is $\mu_\gamma=\hbar\omega_p/\sqrt{1+Q_0(1+d)}.$
Obviously, at other points the condition (\ref{rf}) is not valid,
so that the photons stay ordinary.

Thus, the light bullet is divided into numerous parallel
alternating thin layers of the normal and the Bose-Einstein
condensate photon gas. This state of the photon gas may be called
the intermediate state.

We should emphasize that this phenomenon can also exist in the
case of the usual Compton processes, when the photon-electron
interaction is more important than the photon-photon one. In this
case, in the expression (\ref{quasi}) we should take into account
the variation of density of the electrons.

In conclusion, we have presented a new concept of establishment of
the equilibrium between the photon and the wavepacket of the EM
field (the dense photon bunch). This is the fully relativistic
effect due to the strong EM radiation. We have studied the
possibility of creation of the Bose-Einstein condensates for the
case, when the density of plasma does not change. We have shown
that the inhomogeneous dense photon gas can be found in the
intermediate state. We think that this state is most common in
interaction of the relativistically intense EM waves with plasmas.
 We believe that the intermediate
state can be experimentally observed in atomic gases of alkaline
elements contained in magnetic traps \cite{ant}.

\end{document}